\documentclass[twocolumn,showpacs,amsmath,amssymb,prb,aps,superscriptaddress]{revtex4-1}
\usepackage[dvips]{graphicx}
\usepackage[tight,TABTOPCAP]{subfigure}
\usepackage[]{SIunits}
\usepackage[usenames, dvipsnames]{color}
\usepackage{natbib}

\usepackage{ulem} 
\newcommand\rout{\bgroup\markoverwith
{\textcolor{red}{\rule[.5ex]{2pt}{2pt}}}\ULon}

\newcommand{\ab}{\alpha\beta}

\begin{document}
\title{Finite-size scaling and thermodynamics of model supercooled liquids:\\ Long-range concentration fluctuations and the role of attractive interactions} 
\author{Atreyee Banerjee}
\email{banerjeea@mpip-mainz.mpg.de}
\affiliation{Max Planck Institute for Polymer Research, Ackermannweg 10, 55128 Mainz, Germany}
\author{Mauricio Sevilla}
\affiliation{Max Planck Institute for Polymer Research, Ackermannweg 10, 55128 Mainz, Germany}
\author{Joseph F. Rudzinski}
\affiliation{Max Planck Institute for Polymer Research, Ackermannweg 10, 55128 Mainz, Germany}
\author{Robinson Cortes-Huerto}
\email{corteshu@mpip-mainz.mpg.de}
\affiliation{Max Planck Institute for Polymer Research, Ackermannweg 10, 55128 Mainz, Germany}

\date{\today}

\begin{abstract}
We compute partial structure factors, Kirkwood-Buff integrals (KBIs) and chemical potentials of model supercooled liquids with and without attractive interactions. We aim at investigating whether relatively small differences in the tail of the radial distribution functions result in contrasting thermodynamic properties. Our results suggest that the attractive potential favours the nucleation of long-range structures. Indeed, upon decreasing temperature, Bathia-Thornton structure factors display anomalous behaviour in the $k\to 0$ limit. KBIs extrapolated to the thermodynamic limit confirm this picture, and excess coordination numbers identify the anomaly with long-range concentration fluctuations. By contrast, the purely repulsive system remains perfectly miscible for the same temperature interval and only reveals qualitatively similar concentration fluctuations in the crystalline state. Furthermore, differences in both isothermal compressibilities and chemical potentials show that thermodynamics is not entirely governed by the short-range repulsive part of the interaction potential, emphasising the nonperturbative role of attractive interactions. Finally, at higher density, where both systems display nearly identical dynamical properties and repulsive interactions become dominant, the anomaly disappears, and both systems also exhibit similar thermodynamic properties. 
\end{abstract}

\maketitle
\makeatletter
\let\toc@pre\relax
\let\toc@post\relax
\let\toc@pre\relax
\makeatother

\setlength{\parindent}{0pt}
\section{Introduction}

The supercooled state challenges our understanding of the theory of liquids. In particular, the connection between dynamics, 
which varies considerably upon supercooling, and structure, which appears to remain essentially unchanged, is the subject of 
intense research.~\cite{Filion_NatComm2020,Tanaka_PRL2020,Ciamarra_PRL2020,Landes_PRER2020,Klochko_etal_PRE102_042611_2020,Leocmach_NatComm2012,Berthier_Tarjus_PRL2009}
Model systems with reduced complexity, still retaining essential physical features, provide a direct route to investigate this problem. 
For example, Kob--Andersen mixtures~\cite{kob1995testing} with purely repulsive Weeks-Chandler-Andersen interactions (KAWCA)~\cite{weeks1971role} exhibit substantially
different dynamics compared to their Lennard-Jones counterpart (KALJ).~\cite{kob1995testing} By contrast, their structure,
investigated from the point of view of radial distribution functions, is somewhat similar.~\cite{Berthier_Tarjus_PRL2009, berthier2010critical} \\
The connection between pair correlations and dynamical properties has been extensively investigated.~\cite{gotze1987glass,berthier2010critical}
On the one hand, a variety of studies conclude that two-body contributions are not enough to account for the difference in dynamics between the KAWCA and KALJ 
systems. Perhaps the most well-known example is mode-coupling theory, based on pair correlation functions,
which underestimates these dynamical differences.~\cite{berthier2010critical}
Additionally, deviations in many-body structural descriptors such as triplet~\cite{coslovich2013static} and point-to-set correlations,~\cite{hocky2012growing} 
as well as bond-order distributions~\cite{Toxvaerd_PRE2021} and the packing capabilities of local particle arrangements,~\cite{Tanaka_PRL2020} 
have been observed between the KALJ and KAWCA systems. 
These results indicate that higher-order features may be necessary to resolve the difference in their dynamical properties.~\cite{sciortino2001debye}\\
On the other hand, several studies indicate that two-body structure is enough to describe particular aspects of the dynamics of model supercooled liquids. 
For example, features based on the pair structure have been used to predict diffusion constants from short-time trajectories of the
KALJ model.~\cite{deSouza:2008,Ciamarra:2016,Rudzinski:2019} Concerning the comparison between models, 
Bhattacharyya and coworkers~\cite{banerjee2014role,banerjee2016effect} directly explored structure-dynamics relationships 
in KALJ and KAWCA systems.  In particular, they used the Adam--Gibbs relation,~\cite{adam1965temperature} to connect relaxation time 
to the configurational entropy. Their results demonstrated that the two-body contribution to the entropy plays a significant 
role in distinguishing the dynamics of the two systems.\\
To further contribute to the discussion, recent research efforts have focused on the detailed characterization of the liquid's two-body structure. 
In particular, \textit{softness parameters}, defined via weighted integrals of pair-correlation functions~\cite{Cubuk_PRL2015,Landes_PRER2020} 
or multi-dimensional integrals of partial structure factors,~\cite{nandi2021microscopic} respond to minor structural changes and
can accurately describe dynamical differences. However, either non-trivial reweighting procedures or combinations of local and nonlocal terms
prevents an unambiguous identification of the dominant, short- versus long-range, contributions to the resulting structure-dynamics relationship.\\
The potentially dominant role of short-range pair correlations brings with it yet another dilemma. According to perturbation theory, short-range
repulsive interactions mostly dominate the liquid's structure.~\cite{weeks1971role} By contrast, based on Kirkwood-Buff
theory,~\cite{kirkwood1951statistical} long-range fluctuations in the tail of the pair correlation function have a significant effect 
on the system's solvation thermodynamics.~\cite{Binder-etal-JPhysCondensMatter2-7009-1990, Roman-etal-JChemPhys107-4635-1997, SchnellChemPhysLett504-199-2011,Krueger_JPCLetters2013,Cortes_communication2016} The studies mentioned above investigating KALJ and KAWCA dynamics have mainly focused on short-range contributions. Nevertheless, evidence for the nucleation of long-range structures in glassy systems at low temperatures~\cite{FISCHER1993183,PhysRevE.61.6909, salmon2005topological,Zhang14032} highlights the necessity to carefully address this point. 
Finite-size effects present in computer simulations dramatically affect the tail of the pair correlation function and the $ k \to 0$ 
limit of the structure factor, i.e., the long-range structure properties, which in turn sensitively impact thermodynamic quantities.
Consequently, a careful evaluation of finite-size effects becomes critical for investigating these properties in the supercooled regime.\\ 
In this paper, we investigate various thermodynamic properties of KALJ and KAWCA $\rm{a-b}$ mixtures in the supercooled liquid state.  We calculate structure factors of density, $S_{\rho \rho}(k)$, and concentration, $S_{c c}(k)$, while highlighting the $k \to 0$ limit. The KALJ liquid exhibits anomalous behaviour reflected in a major increase in concentration fluctuations. This anomaly closely resembles the nucleation of nanometric clusters reported by Fischer in low-temperature ortho-terphenyl,~\cite{FISCHER1993183,PhysRevE.61.6909} and it has been recently identified as a general feature present in polydisperse colloidal models. \cite{Klochko_etal_PRE102_042611_2020} By contrast, the purely repulsive KAWCA system remains perfectly miscible in the supercooled state. A finite-size Kirkwood--Buff analysis confirms this picture by enabling the precise identification of the $k\to 0$ limit. Furthermore, we show that the isothermal compressibility and chemical potential of the two models exhibit similar trends with temperature, apart from constant shifts. These differences highlight the nonperturbative role of attractive interactions in the system. To sum up, we demonstrate that seemingly small differences in the tail of the radial distribution function result in significantly different structural and thermodynamic properties for supercooled systems with and without attractive interactions.\\
The paper is organised as follows: we provide the computational details in Sec.~\ref{sec:CompMeth}, present the results in Sec.~\ref{sec:Results} and conclude in Sec.~\ref{sec:Conclusions}.

\section{Computational Details}\label{sec:CompMeth}
\begin{figure*}[t!]
	\includegraphics[scale=0.95]{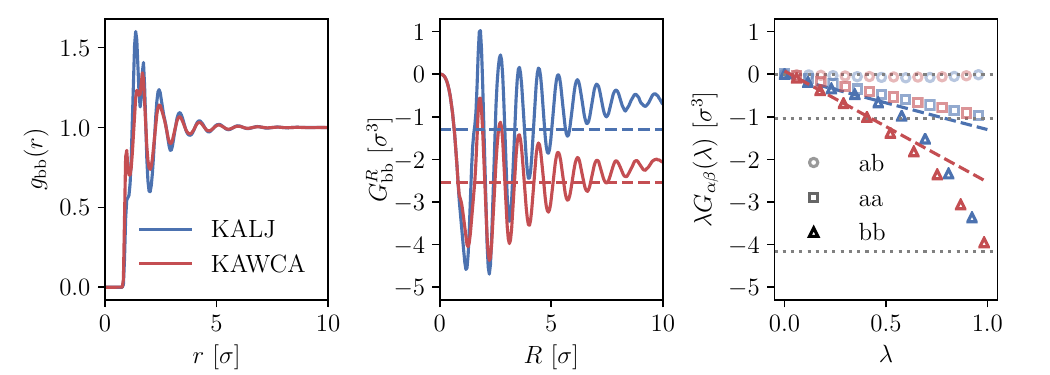}
	\caption{
		Differences between KALJ and KAWCA systems in terms of the $g_{bb}(r)$ component and the KBIs at $T=0.45\epsilon/k_{\rm B}$. (a) Differences between the RDF for the low-concentration $\rm b$-component of the mixture seem to be small and mostly coming from the local structure of the fluid. (b) $G_{{\rm bb}}^R$ as obtained from Eq.~\eqref{eq:KBI_simul} shows a different short-range behaviour and, more importantly, the tails do not converge due to finite-size effects. (c) KBIs obtained using the method described in Ref.~\onlinecite{Cortes_communication2016} (Eq.~\eqref{eq:KBI_fluc}). The KBIs in the thermodynamic limit $G_{\ab}^{\infty}$ are obtained from the slope of a linear fitting of the region $0<\lambda<0.3$. This straight line is indicated for the $\rm{bb}$ case. Horizontal, dark lines correspond to the asymptotic limit $-\delta_{\ab}/\rho_{\alpha}$ with $\delta_{\ab}$ the Kronecker delta and $\rho_{\alpha}$ the number density of the $\alpha$-species. The $G_{\ab}^{\infty}$ values obtained in this way are plotted in panel (b) as horizontal lines.}\label{fig:rdfs_kbis}
\end{figure*}
We have simulated  the Kob--Andersen model, which is a binary mixture (80:20) of Lennard-Jones (KALJ) particles.~\cite{kob1995testing} The inter-atomic pair potential between species $\alpha$ and $\beta$, $U_{\alpha\beta}(r)$, with ${\alpha,\beta}= {\rm a,b}$ is described by a shifted and truncated Lennard--Jones potential\cite{Toxvaerd_jcp2011}, as given by:
\begin{widetext}
\begin{equation}
 U_{\alpha\beta}(r)=
\begin{cases}
 U_{\alpha\beta}^{(LJ)}(r;\sigma_{\alpha\beta},\epsilon_{\alpha\beta})- U_{\alpha\beta}^{(LJ)}(r^{(c)}_{\alpha\beta};\sigma_{\alpha\beta},\epsilon_{\alpha\beta}),    & r\leq r^{(c)}_{\alpha\beta}\\
   0,                                                                                       & r> r^{(c)}_{\alpha\beta}
\end{cases}
\label{LJ_pot}
\end{equation}
\end{widetext}
\noindent where $U_{\alpha\beta}^{(LJ)}(r;\sigma_{\alpha\beta},\epsilon_{\alpha\beta})=4\epsilon_{\alpha\beta}[({\sigma_{\alpha\beta}}/{r})^{12}-({\sigma_{\alpha\beta}}/{r})^{6}]$ and $r^{(c)}_{\alpha\beta}$  is equal to 2.5$\sigma_{\alpha\beta}$ for LJ system and  $r^{(c)}_{\alpha\beta}$  is equal to the position of the minimum of $U_{\alpha\beta}^{(LJ)}$ for the WCA systems (KAWCA).~\cite{weeks1971role} We have added a linear correction so that both the potential and the force go to zero continuously at the cutoff distance.~\cite{Toxvaerd_jcp2011}   
We have used LJ natural units, such that length, temperature and time are measured in $\sigma_{{\rm aa}}$, ${k_{{\rm B}}T}/{\epsilon_{{\rm aa}}}$ and $\tau =\surd({m_{\rm a}\sigma_{{\rm aa}}^2}/{\epsilon_{{\rm aa}}})$, respectively. 
For all the simulations, we have used the following interaction parameters  $\sigma_{\rm aa}$ = 1.0 $\sigma$, $\sigma_{\rm ab}$ = 0.8$\sigma$, $\sigma_{\rm bb}$ =0.88$\sigma$,  $\epsilon_{\rm aa}$ = 1.0 $\epsilon$, $\epsilon_{\rm ab}$ = 1.5$\epsilon$, $\epsilon_{\rm bb}$ = 0.5$\epsilon$, $m_{\rm a}$ = $m_{\rm b}$ = 1.0$m$. 

We have performed two different sets of simulations: the first for the calculation of dynamical and structural properties, and the second for the calculation of chemical potential, which employed a different box geometry and number of particles. 
All simulations have been carried out using the LAMMPS molecular dynamics software~\cite{lammps}. We have performed the first set of simulations in a cubic box with periodic boundary conditions in the canonical ensemble (NVT), using the Nos\'{e}-Hoover thermostat~\cite{evans1985nose} with an integration timestep of 0.005$\tau$ and a time constant of 100 timesteps. The system is composed of $N = 23328$ particles, with $N_{\rm a} = 18664$ particles of type ${\rm a}$.
We have simulated this system at two different densities, $\rho=1.2/\sigma^3$ and $1.6/\sigma^3$ for different temperatures, as specified in the main text.
Starting from the high temperature case, the final configuration of the simulation has been used as an initial configuration for the simulation one (temperature) step below. The same procedure has been followed for the KALJ and KAWCA systems. For all state points, three to five independent simulations with run lengths $> 100\tau_\alpha$ ($\tau_\alpha$ is the $\alpha$-relaxation time estimated from Ref.~\onlinecite{banerjee2016effect}) have been performed. 

To calculate the excess chemical potential, we have used the LAMMPS\cite{lammps} implementation of \texttt{SPARTIAN} already described in Ref. ~\onlinecite{heidari2018spatially}. The {\tt SPARTIAN} method, a variant of the adaptive resolution method~\cite{adress1,*adress2,*adress3,*annurev,*Potestio_Hamiltonian2013,*Potestio_Monte2013}, simulates the coexistence of an atomistic system to its ideal gas representation at a constant density and temperature. We have computed the excess chemical potential of the system as the external potential required to balance the density across the simulation box. To guarantee enough statistics, we have used a slab geometry (An anisotropic box with $L_x=36\sigma$, $L_y=578\sigma$ and $L_z=10\sigma$), also with periodic boundary conditions and at density $\rho=1.2/\sigma^3$, resulting in a system with $N = 250000$ and $N_{\rm a} = 200000$. 
The same protocol as described above has been used to quench the system before performing the \texttt{SPARTIAN} calculation. 
For the \texttt{SPARTIAN} method calculation, we have considered an slab geometry with atomistic region of length of $10\sigma$ and hybrid regions of linear size $10\sigma$ aligned along the $x$ direction.\\
After equilibration, we have performed the \texttt{SPARTIAN} calculations in the canonical ensemble (NVT), using a Langevin thermostat with $dt=0.001\tau$ and damping parameter of $10\tau$. In order to get the correct density profiles and therefore, chemical potential, we have simulated for $3\times 10^6$ simulation steps. 
\section{Results and discussions}\label{sec:Results}
\subsection{Kirkwood Buff analysis}
We consider temperatures in the range $0.45\epsilon/k_{\rm B} \le T \le 6 \epsilon/k_{\rm B}$ for 
KALJ system and $0.3\epsilon/k_{\rm B} \le T \le 6 \epsilon/k_{\rm B}$ for KAWCA system (See Section \ref{sec:CompMeth}).
Visual inspection of the radial distribution functions (RDFs) for both systems reveals that they are almost indistinguishable (Figure S1), and only the RDF $g_{\rm{bb}}(r)$ for the minor component shows relatively small differences, visible at $r<3\sigma$
(Figure \ref{fig:rdfs_kbis}(a)).~\cite{pedersen2010repulsive,banerjee2016effect} However, this direct comparison is misleading: a few
thermodynamic quantities are quite sensitive to small fluctuations in the tail of the RDFs. \\   
One such quantities are the  Kirkwood--Buff integrals (KBIs),~\cite{kirkwood1951statistical} which relate the microscopic structure of 
a liquid mixture to its solvation thermodynamics. For a multi-component system of species $\alpha$ and $\beta$, in equilibrium at temperature $T$, 
the KBIs in the thermodynamic limit (TL) take the form
\begin{equation}
G_{\ab}^{\infty}  = 4\pi \int_0^{\infty} dr\, r^2 (g_{\ab}(r)-1)\, ,
\label{eq:KBI_TL}
\end{equation}
where $g_{\ab}$ is the radial distribution for an infinite, open system. Here, it is obvious from Eq. \ref{eq:KBI_TL} that small deviations for
large $r$ result in important contributions to $G_{\ab}$. In computer simulations, usually far from the thermodynamic limit, Equation \eqref{eq:KBI_TL} 
is often approximated as 
\begin{equation}
G_{\ab}^{R}  = 4\pi \int_0^R dr\, r^2 (g_{\ab}^{\rm c}(r)-1)\, ,
\label{eq:KBI_simul}
\end{equation}
where $g_{\ab}^c(r)$ is the RDF of the closed, finite, system and $R$ is a truncation radius. It is essential to choose $R$ larger 
than the correlation length of the system. Nevertheless, this expression seldom converges due to different finite-size effects. 
Here, it is already clear that $G_{{\rm bb}}^R$ for the 
KALJ and KAWCA systems displays different behaviour (See Figure \ref{fig:rdfs_kbis}(b)).\\
By explicitly including finite-size effects due to the thermodynamic ensemble and the finite integration domains,
we compute the KBIs as~\cite{Cortes_communication2016} 
\begin{equation}
\label{eq:KBI_fluc}
\lambda G_{\ab}(\lambda) 
= \lambda G_{\ab}^{\infty} \left[1-\lambda^{3}\right]  
- \lambda^{4}\frac{\delta_{\ab}}{\rho_{\alpha}} + \frac{c_{\ab}}{V_{0}^{\frac{1}{3}}} \, ,
\end{equation}
where $\lambda \equiv \left(V/V_{0}\right)^{\frac{1}{3}}$ and $G_{\ab}^{\infty}$ is the value of the KBIs in the thermodynamic
limit. We can compute $G_{\ab}(\lambda)$, the KBIs for a subdomain of volume $V$ inside a simulation box of volume $V_0$, 
in terms of fluctuations of the number of particles~\cite{Binder-etal-JPhysCondensMatter2-7009-1990,Roman-etal-JChemPhys107-4635-1997,SchnellChemPhysLett504-199-2011,Krueger_JPCLetters2013,Cortes_communication2016,heidari2018finite,heidari2018fluctuations}\\
\begin{equation}
\label{eq:KBI_fluc2}
G_{\ab}(\lambda) = V\left ( \frac{\langle N_{\alpha}N_{\beta}\rangle' - 
	\langle N_{\alpha}\rangle' \langle N_{\beta}\rangle'}{\langle N_{\alpha}\rangle' \langle N_{\beta}\rangle'} - 
\frac{\delta_{\ab}}{\langle N_{\alpha}\rangle' } \right )\, ,
\end{equation}
where $G_{\ab}(\lambda) \equiv G_{\ab}(V;V_0)$  and the average number of $\alpha$-particles, $\langle N_{\alpha} \rangle'\equiv \langle N_{\alpha}\rangle_{V,V_{0}}$, 
depends on both the subdomain and simulation 
box volumes. Figure \ref{fig:rdfs_kbis}(c) shows the results obtained from Eqs~\eqref{eq:KBI_fluc} and \eqref{eq:KBI_fluc2} 
for the KALJ and KAWCA systems at $T=0.45\epsilon/k_{\rm B}$. These curves are rather similar in both cases,
with a major difference appearing for the $G_{bb}^{\infty}$ component, which can be obtained as the slope of a 
linear fit of $G_{bb}(\lambda)$ within the region $\lambda<0.3$. The resulting values of $G_{{\rm bb}}^{\infty}$ 
are plotted as dashed lines in Figure \ref{fig:rdfs_kbis}(b) to indicate the value at which the KBIs should converge.\\ 
\subsection{Density and concentration structure factors}
\begin{figure}[ht!]
	\subfigure{\includegraphics[]{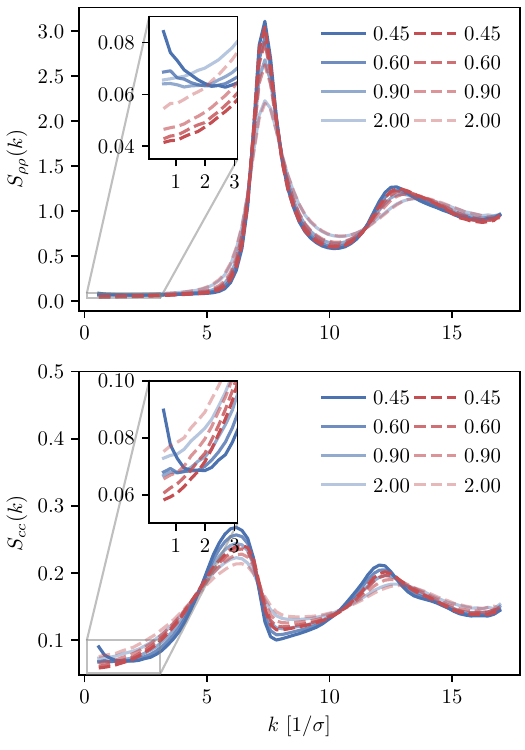}}
	\caption{Density, $S_{\rho\rho}$ (top), and concentration, $S_{cc}$ (bottom), structure factors for both KALJ (blue) and KAWCA (red) systems for the temperatures considered here.}
	\label{fig:btsf}
\end{figure}
As anticipated, fluctuations in the tail of the radial distribution function affect the long-range structure of the fluid.
Hence, to investigate these effects, we compute partial structure factors 
\begin{equation}\label{eq:Sab}
S_{\ab}(k) = x_{\alpha}\delta_{\ab} + 4\pi x_{\alpha}x_{\beta} \rho \int_{0}^{\infty} dr\, r^{2}\frac{\sin k r}{k r} (g_{\ab}(r) - 1)\, ,
\end{equation}
where $k$ is the norm of a reciprocal-lattice vector, $\delta_{\ab}$ is the Kronecker delta, $\rho = \rho_{\rm a} + \rho_{\rm b}$ 
is the total number density and $x_{\alpha} = N_{\alpha}/N$ is the mole fraction of the $\alpha$-species. 
To avoid numerical instabilities at the low $k$ limit~\cite{sedlmeier2011spatial},
we compute the structure factor directly from the simulated trajectory~\cite{berthier2010critical} using the following expression as,
\begin{equation}
 S_{\alpha \beta}(\mathbf{k})=\frac{1}{N}\left\langle \sum_{i\in \alpha}^{N_{\alpha}} \sum_{j\in \beta}^{N_{\beta}} \exp(-i\mathbf{k}\cdot(\mathbf{r}_i-\mathbf{r}_j))\right\rangle,
\label{sq-eqn}
\end{equation}
where $\alpha$ and $\beta$ denote the species, and the indexes $i$ and $j$ run over particles belonging to $\alpha$ and $\beta$, respectively. 
The average runs over the values of $\mathbf{k}$ such that $|\mathbf{k}|=k$ and over the ensemble. 
Partial structure factors are difficult to interpret for liquid mixtures. Hence, we focus on density, $S_{\rho \rho}(k)$,  and concentration, $S_{cc}(k)$, structure factors~\cite{BT_Sofk}
which carry a direct physical meaning.~\cite{Pietro_PCCP2021} 
$S_{\rho \rho}(k)$ and $S_{cc}(k)$ describe the correlation of density and concentration fluctuations in the liquid mixture.
They are defined as
\begin{equation}
\begin{split}
S_{\rho \rho} (k) &= S_{{\rm aa}}(k) + S_{{\rm bb}}(k) + 2S_{{\rm ab}}(k)\, ,\\
S_{c c} (k) &= x_{\rm b}^2 S_{{\rm aa}}(k) + x_{\rm a}^2S_{{\rm bb}}(k) - 2x_{\rm a
}x_{\rm b}S_{{\rm ab}}(k)\, .
\end{split}
\label{eq:BTsq}
\end{equation}
For large $k$-values, the behaviour of $S_{\rho \rho}$ and $S_{cc}$ is rather similar for both systems (See Figure \ref{fig:btsf}). 
	This includes a  first peak at $k_{0}\approx 7.13/\sigma$, followed by a second peak at approximately $1.7 k_{0}$ that develops at low temperatures. 
	This second peak is associated with the nucleation of structural motifs that precede the complete crystallisation of the system. 
	As it has been reported for various metallic glasses, the splitting of this second peak~\cite{KnollSteeb1978}
	results from the optimal facet-sharing configurations of such structural (icosahedral and tetrahedral) motifs 
	that grow upon decreasing temperature.~\cite{VANDEWAAL1995118,Desgranges_PRL2018} In our particular case, we do not observe this feature down to $T=0.45\epsilon/k_{\rm B}$, thus confirming that both systems remain liquid-like.  Concerning the difference between the KALJ and the KAWCA systems, the first and second peaks in the $S_{cc}$ show slightly more structure for the KALJ system at $T=0.45\epsilon/k_{\rm B}$, as expected from the RDF (See Figure \ref{fig:rdfs_kbis}(a)). \\
Perhaps more  interesting, it is apparent from the inset in Figure \ref{fig:btsf} that the KALJ and KAWCA systems show substantially different behaviour in 
	the region of small $k$ (large $r$).  On the one hand, the KAWCA liquid behaves like a normal liquid with monotonically 
	decreasing density fluctuations upon decreasing temperature. On the other hand, the KALJ system exhibits anomalous behaviour, similar to SAXS
	curves obtained for ortho-terphenyl~\cite{PhysRevE.61.6909} and supercooled water,~\cite{kim2017maxima} with clear density
	fluctuations starting around $k\sim 2/\sigma$ ($r\sim 3\sigma$) appearing at temperatures lower than the onset temperature of
	glassy dynamics $T=1\epsilon/k_{\rm B}$ (See Figure S2).~\cite{banerjee2017determination} These results indicate that the two systems display stark structural differences in the supercooled regime,
	with clear long-range density domains ($r>3\sigma$) induced by the presence of attractive interactions in the KALJ mixture.\\
The extrapolation to the $k\to 0$ limit by using Eq.~\eqref{eq:Sab} or ~\eqref{sq-eqn} is not trivial because finite-size effects in the simulation affect the precision in computing structure factors as we approach the linear size of the simulation box.
In the next subsection, we use the relation between the structure factor in the limit $k\to 0$ and the KBIs to investigate this limiting case in more detail. 
\subsection{ KBIs and the $k \to 0$ limit}

Similar to the single component case, the extrapolation to the $k\to 0$ limit provides useful physical information.~\cite{KnollSteeb1978} 
Here,  we use the relation between the structure factor in the limit $k\to 0$ and the KBIs
\begin{equation}
\lim_{k\to 0} S_{\ab} (k) =  x_\alpha \delta_{\ab} + \rho_\alpha x_\beta G_{\ab}^{\infty}\, ,
\end{equation}
thus
\begin{equation}\label{eq:Srr_0}
\begin{split}
\lim_{k\to 0} S_{\rho \rho} (k) &=  \rho_{\rm a} x_{\rm a} G_{{\rm aa}}^{\infty} + \rho_{\rm b} x_{\rm b} G_{{\rm bb}}^{\infty} + 2\rho_{\rm a} x_{\rm b} G_{{\rm ab}}^{\infty} + 1\\
&= \rho k_{\rm B}T\kappa_T + \delta^2 \lim_{k\to 0} S_{cc}(k)\, .
\end{split}
\end{equation}
The last relation in Eq.\eqref{eq:Srr_0} gives two contributions that allows us to connect long-range density fluctuations to both the isothermal compressibility $\kappa_T$ of the system and to concentration fluctuations modulated by the difference in partial molar volumes $v_{\rm a}-v_{\rm b}$, with $\delta = \rho (v_{{\rm a}}-v_{{\rm b}})$.~\cite{KnollSteeb1978} The isothermal compressibility and the partial molar volumes can also be written in terms of the KBIs, namely:
\begin{equation}
\kappa_T =\frac{1+\rho_{\rm a} G_{\rm aa}^{\infty}+\rho_b G_{\rm bb}^{\infty}+\rho_{\rm a} \rho_b(G_{\rm aa}^{\infty}G_{\rm bb}^{\infty}- G_{\rm{ab}}^{\infty\, 2})}{k_B T \eta}\, ,
\label{eq:isot_compress}
\end{equation}
and
\begin{equation}
\begin{split}
v_{\rm a} &= \frac{1+\rho_{{\rm b}}(G_{{\rm bb}}^{\infty}-G_{{\rm ab}}^{\infty})}{\eta}\, ,\\
v_{\rm b} &= \frac{1+\rho_{{\rm a}}(G_{{\rm aa}}^{\infty}-G_{{\rm ab}}^{\infty})}{\eta}\, ,
\end{split}
\end{equation}
where $\eta=\rho_{{\rm a}} + \rho_{{\rm b}} +\rho_{{\rm a}}\rho_{b}(G_{{\rm aa}}^{\infty}+G_{{\rm bb}}^{\infty} -2G_{{\rm ab}}^{\infty})$.

\begin{figure}[ht!]
\subfigure{
\includegraphics[scale=0.95,angle=0]{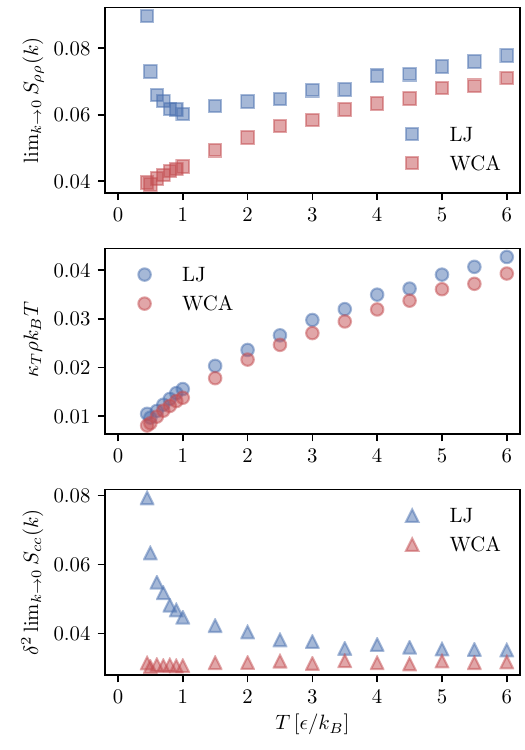}}
\caption{ Density-density correlation function $S_{\rho \rho}(k)$. (Upper panel) $\lim_{k\to 0}S_{\rho \rho}(k)$ obtained 
from the KBIs (Eq.~\eqref{eq:Srr_0}).
At high temperature, both systems present a similar monotonically decreasing behaviour upon decreasing temperature. 
At the onset temperature of glassy dynamics ($T=1\epsilon/k_{\rm B}$),~\cite{banerjee2017determination} the data for the KALJ system shows 
an inflexion point which signals density-density correlations visible for distances longer than $r=2.5\sigma$.
Individual components of $\lim_{k\to 0}S_{\rho\rho}(k)$:  (Middle panel) $\kappa_T \rho K_{\rm B}T$ and  (Lower Panel)
$\delta^2 \lim_{k\to 0}S_{cc}(k)$ with $\delta = \rho(v_{{\rm a}}-v_{{\rm b}})$ the product of the total density with the
difference in partial molar volumes. It is apparent that the contrast in $S_{\rho\rho}$ originates from major concentration fluctuations
present in the KALJ system, as indicated by $S_{cc}(k)$.
}
\label{fig:Srr}
\end{figure}
We use the definition in Eqs~\eqref{sq-eqn} and \eqref{eq:BTsq} to compute $S_{\rho \rho}(k)$,
and compare with the $\lim_{k\to 0}S_{\rho \rho}(k)$ obtained from the KBIs (Eq.~\eqref{eq:Srr_0}). 
The results, presented in Figure \ref{fig:Srr} (Top panel),  confirm the information given by the partial structure factors (See Figure S2  for a comparison between the values obtained from the structure factor and the KBIs). Namely, in contrast to the KAWCA system, the KALJ system exhibits increasingly large density fluctuations upon decreasing temperature. To investigate the origin of the anomaly, we investigate the contributions to $S_{\rho  \rho}$ separately as given by the r.h.s. of Eq.~\eqref{eq:Srr_0}. The middle and lower panels of Figure \ref{fig:Srr} splits $S_{\rho \rho}$ into isothermal compressibility and concentration fluctuation terms, respectively. There, it is apparent that the anomalous behaviour exhibited by the KALJ system at low $k$ values is due to the formation of long-range concentration domains (red and blue triangles). By contrast, the isothermal compressibility contribution remains nearly the same for both systems (red and blue circles). \\
The anomaly observed in the limit $k\to 0$ in Figure ~\ref{fig:btsf} has also been reported in Ref.~\onlinecite{Ingebrigtsen_PRX2019}. The authors suggest a plausible explanation involving the gas-liquid phase separation of the KALJ system. However, we note that the gas-liquid coexistence region for the KALJ system is still far from the point $\rho = 1.2/\sigma^3$, $T=0.45\epsilon/k_{\rm B}$~\cite{sastry2000liquid, testard2011influence} (See also the discussion in Sec. IV in Ref.~\onlinecite{Ingebrigtsen_PRX2019}). Moreover, we observe the non-monotonic behaviour starting just below the onset temperature of glassy dynamics  $T=1\epsilon/k_{\rm B}$, which is even farther away from the coexistence region. Moreover, the virial part of the pressure remains positive even at the lowest temperature considered for the KALJ model (See Fig. 1 in Ref.~\onlinecite{berthier2011role} and Figure S3). For the KAWCA model, as expected, the system's pressure is systematically higher than the KALJ pressure due to the absence of attractive interactions. In general, the positive pressure of our simulated state points already suggests that the anomalous behaviour is not due to gas-liquid coexistence.\\
\begin{figure}[h]
	\includegraphics[scale=0.95,angle=0]{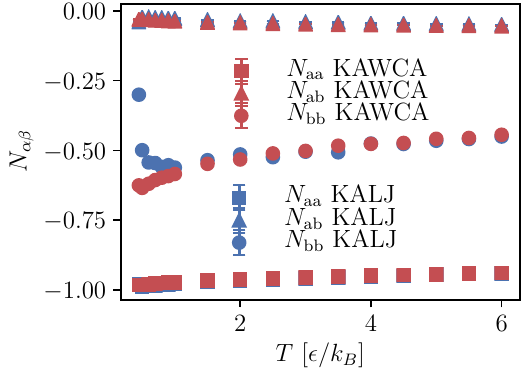}
	\caption{
		Excess coordination number ($N_{\ab} = \rho_{\beta}G_{\ab}^{\infty}$) as a function of temperature for both KALJ and KAWCA systems. $N_{\rm ab}$ close to zero corresponds to a preferential ${\rm a-b}$ effective interaction. Below the onset temperature of glassy dynamics upon cooling,  $N_{\rm bb}$ gets close to zero for the KALJ system, indicating a growing preferential ${\rm b-b}$ effective interaction, ultimately leading to phase segregation.
	}\label{fig:SI_coornumx}
\end{figure}
Plots of the excess coordination number ($N_{\ab} = \rho_{\beta}G_{\ab}^{\infty}$) as a function of temperature provide a clear insight (See Figure \ref{fig:SI_coornumx}).  
As expected from the model, the effective interaction between $\rm a$ and $\rm b$ particles is favoured in both systems at all temperatures: excess coordination numbers are close to zero. Below the onset temperature of glassy dynamics, 
the excess coordination number shows a collective tendency for the KALJ mixtures to increase ${\rm b-b}$ effective interactions upon cooling. 
This propensity is not observed in the KAWCA case. We underline here that these concentration domains for the KALJ system resemble the behaviour
discovered by Fischer~\cite{FISCHER1993183} for supercooled ortho-terphenyl. Namely, anomalies in the structure factor at low $k$-values,
which are not commensurate with the isothermal compressibility, are connected to the nucleation of nanometric structures.~\cite{Stevenson_JPCA2011} Furthermore, our results agree with recent theoretical efforts demonstrating that the low $k$ portion of the structure factor  for polydisperse colloidal systems can be separated into a compressibility contribution and a term related to composition fluctuations.~\cite{Klochko_etal_PRE102_042611_2020}\\
\begin{figure}[h]
	\includegraphics[scale=0.95,angle=0]{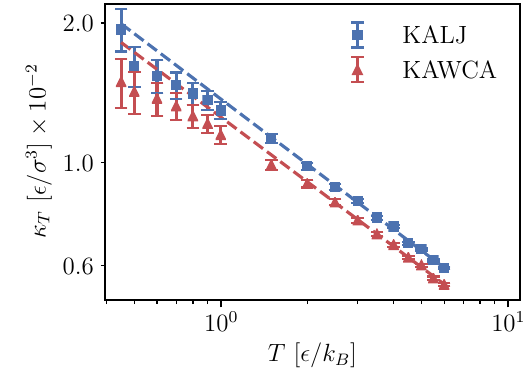}
	\caption{
		Bulk isothermal compressibility $\kappa_T$, calculated from Eq.~\eqref{eq:isot_compress},
		as a function of temperature for KALJ and KAWCA systems (Log-log representation). We observe that 
		a power law relationship holds as $\kappa_T = \kappa^0_T T^{-\gamma}$ with $\gamma = 0.46 \pm 0.01$ and $0.45\pm 0.01$ for KALJ and KAWCA, respectively. The dashed lines are the corresponding power law fitting.}\label{fig:iso_com}
\end{figure}
We now focus on the isothermal compressibility (Eq. \ref{eq:isot_compress}). In Figure \ref{fig:iso_com}, we present
a log-log plot of $\kappa_T$ vs $T$, where it is apparent that the KALJ system is systematically more compressible than the KAWCA system at all
temperatures considered here. Hence, it is again clear that small differences in the tail of the RDFs result in sizeable differences in their
thermodynamic properties. Furthermore, a power-law behaviour 
$\kappa_T = \kappa_T^0 T^{-\gamma}$ is apparent
with $\gamma = 0.46 \pm 0.01$ for the KALJ system and $\gamma = 0.45\pm 0.01$ for the KAWCA system.
Below the onset temperature of glassy dynamics, both systems deviate from this power law and become comparatively less compressible in the deeply supercooled region. One would expect that, for a system undergoing a gas-liquid separation, compressibility substantially increases upon approaching the gas-liquid region. By contrast, the behaviour observed in Figure~\ref{fig:iso_com} suggests that this is not the case. 
Finally, the existence of this power law, including the low-temperature deviations~\cite{kim2017maxima},
is somewhat similar ($\gamma = 0.40 \pm 0.01$)~\cite{Spaeh_Apparent2019} to the one observed experimentally in supercooled water.
\subsection{Chemical potential}
\begin{figure}[ht]
	\includegraphics[scale=0.95,angle=0]{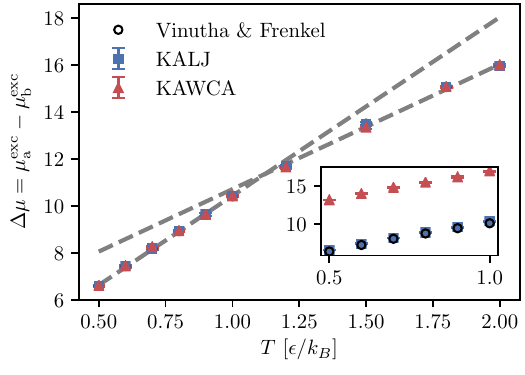}
	\caption{ Difference of excess chemical potentials between species a and b for both, KALJ and KAWCA systems. The KAWCA system results were shifted by a constant in order to mach the lowest temperature $T=0.5\epsilon/k_{\rm B}$, indicating that the potential energy can be approximated to $U_{\rm{LJ}} \approx U_{\rm{WCA}} + U_{\rm{Attractive}}$. A change in the behaviour with $T$, indicated by the dashed-grey lines, is apparent at the onset temperature of glassy dynamics. The inner plot shows the difference of chemical potential for KALJ and KAWCA (without shifting). Results for the KALJ system in the temperature range 0.5 -- 1.0$\epsilon/k_{\rm B}$ well compare with results available in the literature.~\cite{vinutha2021computation}}
	\label{fig:chem_pot}
\end{figure}

Finally, we compute the excess chemical potential for both systems (Figure \ref{fig:chem_pot}) 
using the {\tt SPARTIAN} method.~\cite{heidari2018spatially} Recent calculations of the chemical potential for the KALJ system
in the range of temperature $0.5\epsilon/k_{\rm B} < T < 1.0\epsilon/k_{\rm B}$ are in excellent agreement with our results.~\cite{vinutha2021computation} 
At the onset temperature of glassy dynamics, there is a transition between two regimes, reflecting the tendency for the system to minimise its free energy. 
The fact that the curves for the KALJ and the KAWCA systems are identical up to a constant factor is a consequence of writing the LJ potential 
energy as  $U_{\rm{LJ}} \approx U_{\rm{WCA}} + U_{\rm{Attractive}}$. This expression lies at the foundation of perturbation theory 
that assumes that $U_{\rm Attractive}$ is very small compared to $U_{\rm WCA}$. However, the sizeable difference in chemical potential
($\approx 5 \epsilon$) indicates that this approximation does not hold in this case.\\
Similarly to other thermodynamic properties like excess~\cite{banerjee2016effect} and configurational entropy~\cite{banerjee2014role}, isothermal compressibility and chemical potential results confirm that perturbation theory is not valid for the KALJ and KAWCA systems at $\rho=1.2/\sigma^3$ since attractive interactions induce 
non-perturbative structural effects. In the following section, we investigate these systems at higher density, namely $\rho=1.6/\sigma^3$, where we expect that repulsive interactions play an increasingly dominant role.~\cite{Berthier_Tarjus_PRL2009,berthier2010critical} 
\subsection{KALJ and KAWCA mixtures at $\rho = 1.6/\sigma^3$}\label{sec:SI_highrho}
\begin{figure}[ht!]
	\subfigure{\includegraphics[]{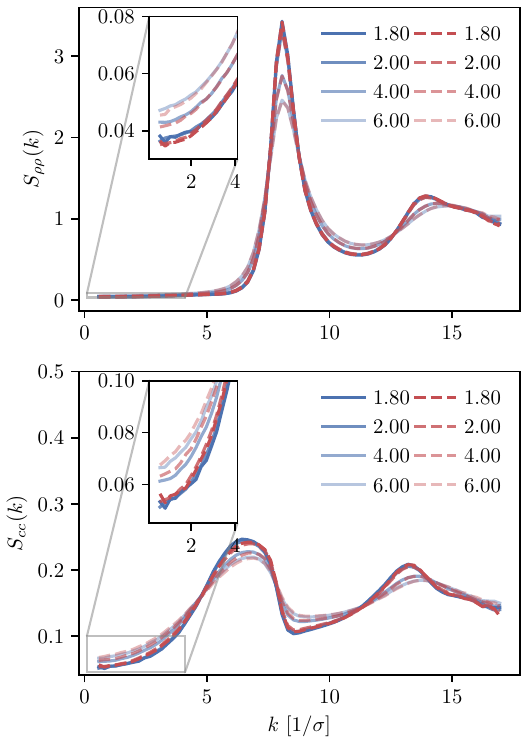}}
	\caption{
		Density and concentration structure factors, $S_{\rho\rho}$ and $S_{cc}$, for the KALJ and KAWCA systems at a higher density ($\rho =1.6$), in the range of temperature considered here.
	}\label{fig:SI_btsf_1.6}
\end{figure}
\begin{figure}[h]
	\includegraphics[]{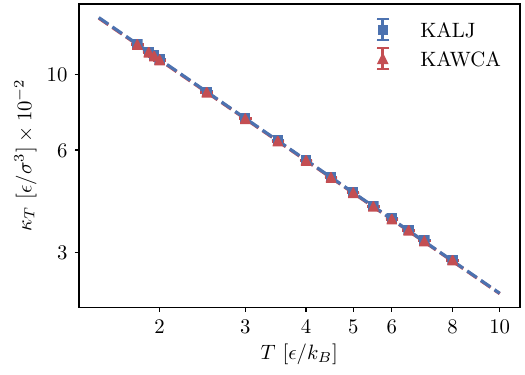}
	\caption{Bulk isothermal compressibility $\kappa_T$ for the KALJ and KAWCA systems at a higher density ($\rho=1.6/\sigma^3$)}\label{fig:SI_comp_1.6}
\end{figure}
We perform a similar thermodynamic analysis for KALJ and KAWCA systems at $\rho=1.6/\sigma^3$. Our results show that density and concentration structure factors (Figure \ref{fig:SI_btsf_1.6}) are nearly identical for both systems in the range of temperature considered. Structure factors in the limit $k\to 0$, in particular, show no evidence for the nucleation of long-range structures.  Dynamical properties for these mixtures available in the literature~\cite{berthier2011role,banerjee2016effect,banerjee2020fragility} reveal 
that both systems exhibit similar structural and dynamical properties at this density. Therefore, we conclude that long-range concentration fluctuations might be closely connected to the significant mismatch between dynamical properties of the two systems at $\rho = 1.2/\sigma^3$. \\
Concerning the isothermal compressibility (Figure \ref{fig:SI_comp_1.6}), the two systems are essentially indistinguishable in the whole temperature range. As a reference, the onset temperature of glassy dynamics for KALJ and KAWCA systems at this density is close to $2.80\epsilon/k_B$.~\cite{banerjee2017determination}
The two systems hence behave similarly well below the onset temperature of glassy dynamics.  This result highlights the dominant role played by attractive interactions in determining thermodynamic properties of high-density liquids.
\subsection{Crystallisation of the KAWCA system}
\begin{figure}[h]
			\subfigure{\includegraphics[width=0.4\textwidth]{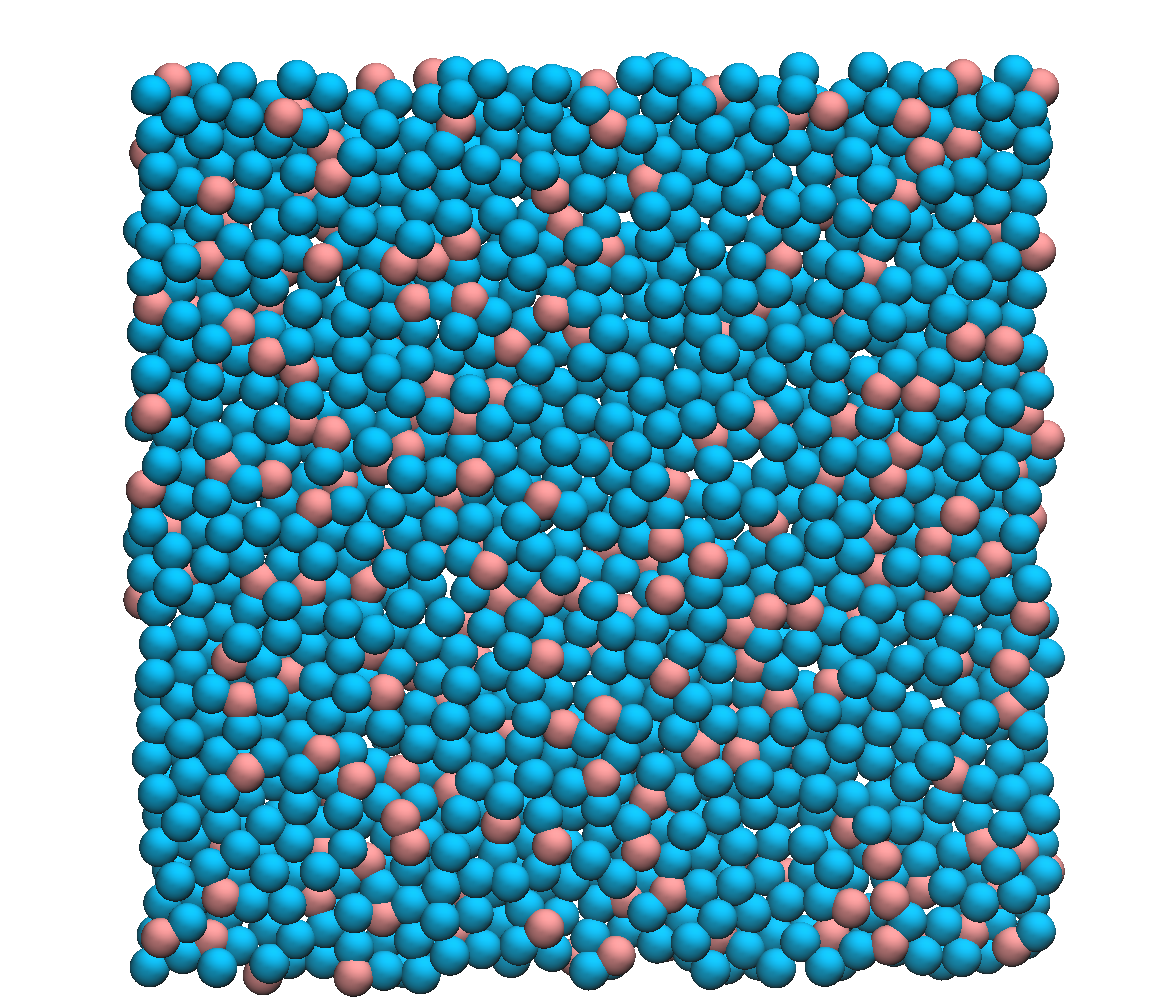}}
	\subfigure{\includegraphics[width=0.4\textwidth]{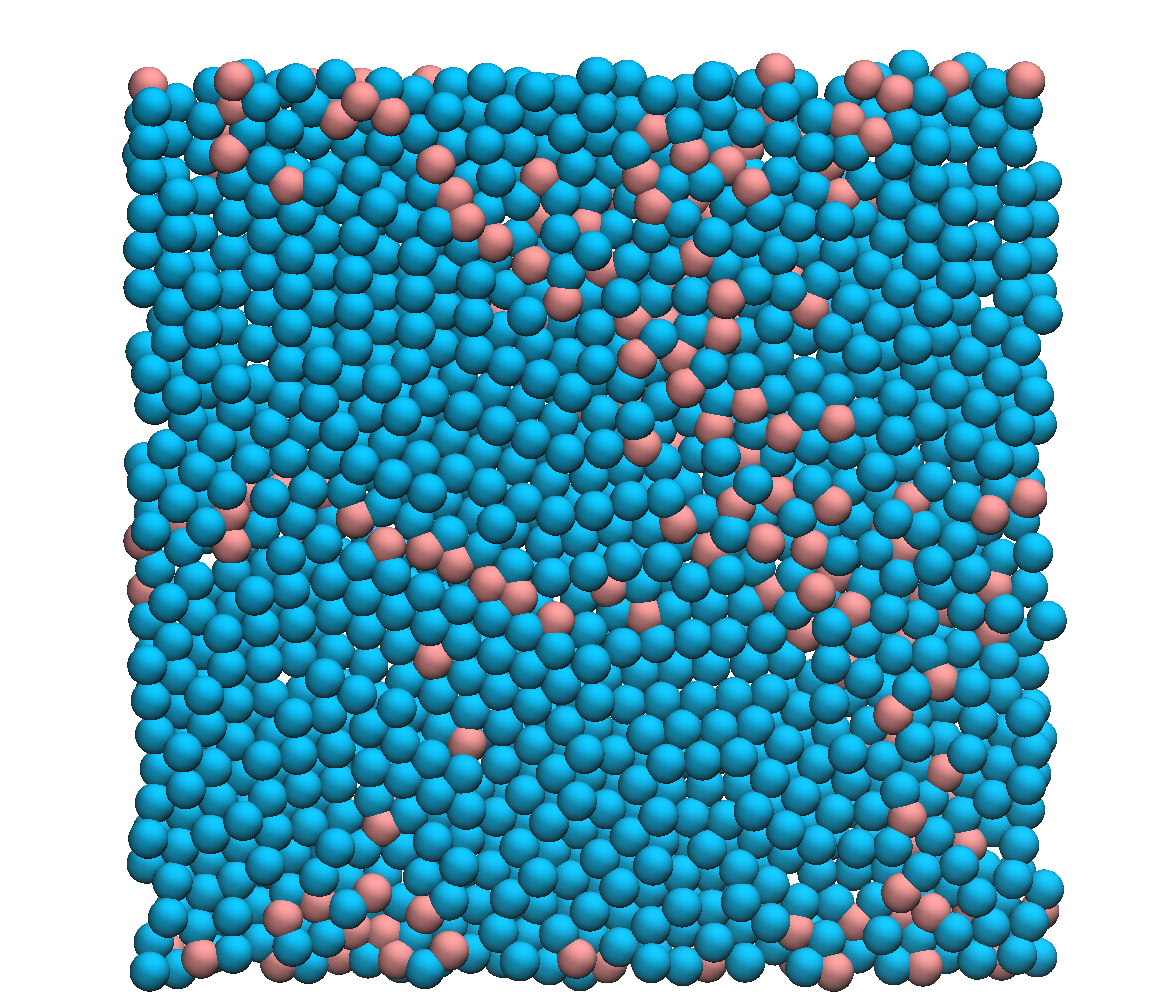}}
	
	\caption{Snapshot of the KAWCA system at $T=0.45\epsilon/k_{\rm B}$ (Top) and $T=0.35\epsilon/k_{\rm B}$ (Bottom).
	}\label{fig:snap}
\end{figure}

\begin{figure}[ht!]
	\subfigure{\includegraphics[]{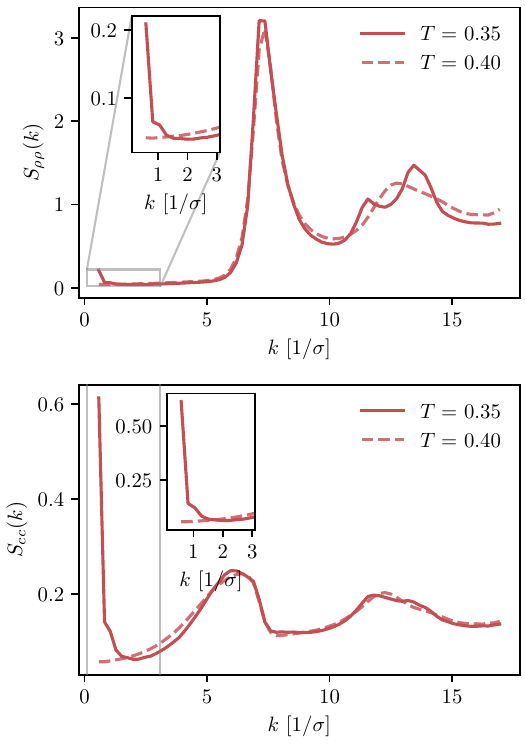}}
	\caption{
		Density and concentration structure factors, $S_{\rho\rho}$ and $S_{cc}$, for the KAWCA system in the temperature range $0.3\epsilon/k_{\rm B} < T < 0.45\epsilon/k_{\rm B}$.
	}\label{fig:SI_btkawcalowt}
\end{figure}

In the last section, we investigate the crystallisation of the KAWCA system at $\rho = 1.2/\sigma^3$. We further decrease the temperature down to $T=0.35\epsilon/k_{\rm B}$.  Figure \ref{fig:snap} shows snapshots of the system at $T=0.45\epsilon/k_{\rm B}$ (Left panel) and $T=0.35\epsilon/k_{\rm B}$ (Right panel). It is apparent from the figure that the system at $T=0.45\epsilon/k_{\rm B}$ appears like a miscible liquid. Conversely, the system at $T=0.35\epsilon/k_{\rm B}$ shows crystalline domains with a clear tendency for phase-segregation. 
Density and concentration structure factors (Figure \ref{fig:SI_btkawcalowt}) enable us to validate this crystallisation-demixing scenario. 
In particular, we  observe  the splitting of the second peak of $S_{\rho \rho}(k)$ at $1.7  k_0$ with  $k_0\approx 7.13/\sigma$ that indicates the presence of facet-sharing domains between neighbouring crystalline regions. Perhaps more interesting, the structure factors at $k<2/\sigma$ show a marked formation of long-range domains qualitatively similar to the ones present for the  KALJ case below onset temperature. \\
 These results indicate that the crystallisation of the KAWCA system is a process driven by phase segregation.
We note that similar behaviour to that observed for the low-temperature density and concentration structure factors for KAWCA 
has also been observed in polydisperse glass-forming systems.~\cite{Ninarello_PhysRevX2017} Furthermore, recent GPU simulations report the
crystallisation of the KALJ system~\cite{Ingebrigtsen_PRX2019} due to composition fluctuations similar to the ones investigated in this work. 
Indeed, our results also support previous claims pointing out demixing as a precursor for crystalisation in the modified KA model~\cite{nandi2016composition} and in liquid metals.~\cite{desgranges2014unraveling}\\
We conclude here that the presence of long-range concentration fluctuations is a qualitatively common feature for both KALJ and KAWCA systems. Perhaps more important, it is not directly related to gas-liquid coexistence present in the KALJ system. Indeed,  in the crystalline state ($T=0.35\epsilon / k_{\rm B}$), the KAWCA system exhibits a significant growth of concentration fluctuations, apparent in the lower panel in Figure \ref{fig:SI_btkawcalowt}. The attractive interactions present in the KALJ system favour the nucleation of concentration fluctuations starting at relatively high temperatures ($T=1.00\epsilon/k_{\rm B}$). Although the KAWCA system crystallises in our simulation timescale as its dynamics are considerably faster than for the KALJ system, our results suggest that both systems crystallise upon demixing by following a similar pathway.\\ 
\section{Conclusions}\label{sec:Conclusions}
We compute various thermodynamic properties of model supercooled liquids, with (KALJ) and without (KAWCA) attractive interactions at density $\rho=1.2/\sigma^3$. We aim at studying whether fluctuations in the tail of the two-body correlation function induce significant thermodynamic differences between the two systems. Density and concentration structure factors in the limit $k\to 0$ indicate that the KALJ system exhibits anomalous structural behaviour that we identify as the nucleation of long-range concentration domains. Conversely, the KAWCA system behaves like a normal liquid, with density and concentration structure factors decreasing monotonically. A finite-size Kirkwood-Buff analysis used to extrapolate to the $k\to 0$ limit confirms this picture. Differences in isothermal compressibilities and chemical potentials highlight the non-perturbative role of attractive interactions. Results of the crystallisation of the KAWCA system suggest that the anomaly, enhanced by the presence of attractive interactions, is a common feature of both models. All our results indicate that these long-range concentration fluctuations are not connected to gas-liquid coexistence, implying that demixing precedes crystallisation in both systems.  Finally, upon increasing density ($\rho = 1.6/\sigma^3$), where KALJ and KAWCA systems show similar dynamical properties, the KALJ anomaly disappears, and both systems exhibit nearly identical thermodynamic properties. Hence, we speculate that there might be a connection between large-scale concentration fluctuations and the significant dynamical slow down of the KALJ system in the deeply supercooled regime.
\begin{acknowledgments}
The authors thank Kurt Kremer for his insightful discussions and his critical reading of the manuscript. They are also grateful to Pietro Ballone, Burkhard D\"unweg, Smarajit Karmakar and Werner Steffen for their valuable feedback and suggestions. R.C.-H. thankfully acknowledge funding from SFB-TRR146 of the German Research Foundation (DFG).
 \end{acknowledgments}
A.B. and M.S. contributed equally to this work. 
\bibliography{refs} 
\end{document}